\documentclass[a4paper,12pt]{article}
\usepackage{mathrsfs}
\usepackage{amsfonts}
\usepackage{amssymb}
\usepackage{amsthm}
\usepackage{amsmath}
\usepackage{authblk}
\usepackage{url}
\allowdisplaybreaks

\newtheorem{Theorem}{\sc Theorem}

\newtheorem{Lemma}[Theorem]{\sc Lemma}

\newtheorem{Remark}[Theorem]{\sc Remark}

\newtheorem{Problem}[Theorem]{\sc Problem}

\def\sqr#1#2{{
    \vcenter{
         \vbox{\hrule height.#2pt
               \hbox{\vrule width.#2pt height#1pt \kern#1pt
                     \vrule width.#2pt
               }
               \hrule height.#2pt
         }
    }
}}
\def\square{\mathchoice\sqr84\sqr84\sqr{2.1}3\sqr{1.5}3}

\def\div{\mathop{\rm div}\nolimits}

\def\bar{\overline}
\def\real{\mathbb{R}}

\newcommand{\R}{{\if mm {\rm I}\mkern -3mu{\rm R}\else \leavevmode
\hbox{I}\kern -.17em\hbox{R} \fi}}

\def\lista#1
{{ \itemindent 0.0cm \labelsep .2cm \leftmargin 0.8cm \rightmargin
0.0cm \labelwidth 0.6cm \topsep 0.0mm
\parsep 0.0mm
\itemsep 0.0mm
\begin{list}{}
{ \setlength{\leftmargin}{1.0cm} \setlength{\rightmargin}{0.0cm}
\setlength{\parsep}{.0mm} \setlength{\topsep}{.0mm}
\setlength{\parskip}{.0cm} \setlength{\itemsep}{.0cm} }
{#1}\end{list}} }

\begin{document}

\title{ 
Evolution Boussinesq model with nonmonotone friction and heat flux boundary conditions
\thanks{\ Research supported by the Marie Curie 
International Research Staff Exchange
Scheme Fellowship within the 7th European Community Framework
Programme under Grant Agreement No.\  295118 and the National
Science Center of Poland under the Maestro Advanced Project 
No.\ DEC-2012/06/A/ST1/00262. 
}}

\author{Pawel Szafraniec\footnote{E-mail:
pawel.szafraniec.wmii@gmail.com.}}

\affil{Faculty of Mathematics and Computer Science \\
Jagiellonian University in Krakow \\
ul. Lojasiewicza 6, 30-348 Krakow, Poland}

\date{}

\maketitle

\bigskip

\noindent {\bf Abstract.} \ 
In this paper we prove the existence and regularity of a solution to a two-dimensional system of evolutionary hemivariational inequalities which describes the Boussinesq model with nonmonotone friction and heat flux. We use the time retardation and regularization technique, combined with a regularized Galerkin method, and recent results from the theory of hemivariational inequalities.  

\bigskip


\vskip 4mm


\thispagestyle{empty}

\newpage


\section{Introduction}\label{Introduction}
The Boussinesq system of hydrodynamics equations arises from several physical problems when the fluid varies in temperature from one place to another, and we simultaneously observe the flow of fluid and heat transfer. The system couples incompressible Navier--Stokes equations for the fluid velocity and the thermodynamic equation for the temperature distribution. For the derivation of the Boussineq equations, see~\cite{BOUSS, DUT, MILH}. The mathematical theory of Navier--Stokes equations has been of the strong interest in the mathematical commmunity for many years, the basic references are monographs \cite{TEMAN1} and \cite{TEMAM2}. The coupled system of incompressible Navier--Stokes problem with the heat equation has been studied in both static and evolutionary cases by many authors, e.g. see~\cite{DIAZ, KOMO, LEE, MIAO} and the references therein.

In the present paper we study the Boussinesq system which consists with two evolutionary partial differential equations of parabolic type. We impose mixed nonmonotone subdifferential boundary conditions. More precisely, we divide the boundary into two parts. On one part the usual Dirichlet condition applies. On the other part, called the contact boundary, we consider a nonmonotone friction law for the velocity, as well as a nonmonotone law for the heat flux. Because of these  nonmonotone conditions, the formulation of the problem based on a variational inequality approach and the notion of convex subdifferential can not be applied. It is worth noting that the subdifferential boundary conditons for Navier--Stokes equations in the convex case have been studied in~\cite{Fujita,Fujita2} and more recently in \cite{KASHI}. These authors consider multivalued boundary conditions generated by the subdifferential of the norm function. Our work generalizes some of the aformentioned results to the problems with boundary conditions described by the Clarke subdifferential of locally Lipschitz functions, cf.~\cite{CLARKE}. For this reason, we use the theory of hemivariational inequalities to derive the weak formulation of the problem. For the mathematical theory of hemivariational inequalities modeling  stationary (time--independent) problems, we refer to~\cite{GOMON, HUANG, NP, PANA} and the references therein. For the evolutionary hemivariational inequalities and their various applications to mechanics, we refer the reader to~\cite{MIGPARA, MIGP,MIGUNI} and the recent monograph~\cite{MOSBOOK}.

Furthermore, in our approach we introduce a strong coupling between the Navier--Stokes equations and the heat equation. This coupling together with an additional presence of nonmonotone contact conditions represents the main difficulty of the system under consideration. 
Note that the Navier--Stokes equations and Stokes problems with nonmonotone boundary conditions and without such coupling have been studied in~\cite{OCHAL2, OCHAL1} and \cite{SLU}. 
Finally, we mention that the main tools in the present paper are abstract results from the theory of hemivariational inequalities, cf.~\cite{MIGPARA} and the time retardation method. The latter technique has been successfully applied to coupled systems in viscoelastic damage, as well as to Stefan problem and thermistors. The time retardation method allowed to obtain interesting existence results in~\cite{FRESHI,KUTSHIL} and~\cite{SHILL1}.

The structure of this paper is as follows. In Section~\ref{Prelim}, we present the preliminary material used later. Section~\ref{Statement} describes the physical setting and the classical formulation of the Boussinesq problem. Section~\ref{SECSLIP} contains the variational formulation of the problem and the proof of our main result on existence and regularity of the solution.


\section{Preliminaries}\label{Prelim}

In this section we recall definitions and notations use throughout the paper. 

We first recall the definitions of the generalized directional 
derivative and the generalized gradient of Clarke for a locally
Lipschitz function $\varphi \colon X \to \real$, where $(X,\|\cdot\|_X)$ is a Banach
space (see~\cite{CLARKE}). 
The generalized directional derivative of $\varphi$ at $x \in X$ in the
direction $v \in X$, denoted by $\varphi^{0}(x; v)$, is defined by
$$\displaystyle 
\varphi^{0}(x; v) =
\limsup_{y \to x, \ t\downarrow 0}
\frac{\varphi(y+tv) - \varphi(y)}{t}.
$$
The generalized gradient of $\varphi$ at $x$, denoted by
$\partial \varphi(x)$, is a subset of a dual space $X^*$ given by
$\partial \varphi(x) = \{ \zeta \in X^* \mid \varphi^{0}(x; v) \ge 
{\langle \zeta, v \rangle}_{X^* \times X}$  
for all $v \in X \}$. 

Let $\Omega$ be a bounded domain in $\mathbb{R}^2$ 
with smooth boundary $\Gamma$ consisting of two open disjoint sets 
$\Gamma_1$ and $\Gamma_0$ such that 
$\Gamma = \bar{\Gamma_0} \cup \bar{\Gamma_1}$.
For a vector $\xi \in \mathbb{R}^2$, we denote by 
$\xi_\nu$ and $\xi_\tau$ its normal and tangential components on the boundary, 
i.e., $\xi_\nu = \xi \cdot \nu$ and $\xi_\tau = \xi - \xi_\nu \nu$, 
where dot denotes the inner product in $\real^2$ 
and $\nu$ is the outward unit normal vector to $\Gamma$.

\medskip

We introduce the following function spaces
\begin{eqnarray*}
&
E_0 = \{v \in H^1(\Omega)^2 \mid v=0 \ \mbox{on} \ \Gamma_0, ~v_\nu =0 \ \mbox{on} \ \Gamma_1 \}, \quad 
V = \{ \theta \in H^1(\Omega) \mid \theta=0 \ \mbox{on} \ \Gamma_0     \}.
\end{eqnarray*}

\noindent
Moreover, we introduce divergence--free spaces
$H_{\sigma}^1 (\Omega)^2 = \{v \in H^1(\Omega)^2  \mid \div v=0 \}$ and  
\begin{equation*}
E = E_0 \cap H_\sigma^1(\Omega)^2.
\end{equation*}

\noindent 
For a finite time interval $(0, T)$, we define the following spaces
\[
\mathbb{E} = \{\, v\in L^2(0,T;E) \mid \dot{v} \in L^2(0,T;E^*)\, \}
\]
and
$$
\mathcal{W} = \{\, v\in L^2(0,T;V) 
\mid \dot{v} \in L^2(0,T;V^*)\, \}.
$$ 
For convenience we denote $\mathcal{E}= L^2(0,T;E)$ and $\mathcal{V}=L^2(0,T,V)$.
The space $H$ is defined as the closure of $\{v \in C_0^\infty({\bar{\Omega}})^2 \mid \mbox{div} \, v = 0, v=0 \ \mbox{on} \ \Gamma_0, v_\nu =0 \ \mbox{on} \ \Gamma_1 \}$ in the $L^2(\Omega)^2$ norm. By $\gamma_s$ and $\gamma$ we denote the trace operators $\gamma_s\colon E\to L^2(\Gamma)^2$ and $\gamma\colon V\to L^2(\Gamma)$.

\medskip

We define bilinear and trilinear forms $a_0 \colon H^1(\Omega)^2 \times H^1(\Omega)^2 \to \real$ , $a_1 \colon H^1(\Omega)^2 \times H^1(\Omega)^2 \times H^1(\Omega)^2 \to \real$, $b_1 \colon H^1(\Omega)^2 \times H^1(\Omega) \times H^1(\Omega) \to \real$ and $c \colon H^1(\Omega)^2 \times L^2(\Omega) \to \real$ by
$$
a_0(u,v)= \frac{\alpha}{2} \sum_{i,j=1}^2 \int_{\Omega} 
\left( \frac{ \partial u_i}{\partial x_j} + \frac{ \partial u_j}{\partial x_i}\right) 
\left( \frac{ \partial v_i}{\partial x_j} + \frac{ \partial v_j}{\partial x_i}\right) \, dx 
\ \ \mbox{for} \ \ u, v \in H^1(\Omega)^2,
$$
where $\alpha > 0$,
$$ 
a_1(u,v,w) = \int_{\Omega} ((u\cdot \nabla)v) \cdot w \, dx 
\ \ \mbox{for} \ \ u, v, w \in H^1(\Omega)^2,
$$
\[
b_1(v,\eta,\zeta) = \int_{\Omega} v \cdot \nabla \eta \, \zeta \, dx \ \ \mbox{for} \ \ v\in H^1(\Omega)^2, \eta,\zeta \in H^1(\Omega),
\]
$$
c(v,q)=-\int_{\Omega} (\mbox{div}\, v) \, q \,dx  
\ \ \mbox{for} \ \ v \in H^1(\Omega)^2, \ q\in L^2(\Omega).
$$
\noindent
We introduce the following functional $b_0 \colon H^1(\Omega) \times H^1(\Omega) \times H^1(\Omega) \to \real$ 
defined by
\[
b_0(\mu,\eta,\zeta) =  \int_{\Omega}  k(\mu)\nabla \eta \cdot \nabla \zeta\, dx \ \ \mbox{for} \ \ \mu,\eta,\zeta \in H^1(\Omega),
\]
where $k \colon \real\to \mathbb{R}$. We also define the operators $A_0$, $A_1 \colon E \to E^*$ , $B_0 \colon V\times V\to V^*$, $B_1 \colon E\times V \to V^*$ by
\begin{equation}\label{oper}
\langle A_0 u, v \rangle = a_0(u,v), \ \ \ \langle A_1 u,v \rangle = a_1(u,u,v) 
\ \ \mbox{for} \ \ u, v \in E,
\end{equation}
\begin{equation}
\langle B_0(\mu, \eta),\zeta\rangle = b_0(\mu, \eta,\zeta), \ \ \ \langle B_1(u,\eta) , \zeta \rangle = b_1(u,\eta,\zeta) \ \ \mbox{for} \ u\in E, \ \eta,\zeta \in V,
\end{equation}
respectively. 

In what follows, we recall the properties of the forms $a_1$ and $b_1$. 
The proof of the following lemma can be found, for example, in Lemmata 3.4 and 3.5 in~\cite{TEMAN1} and Chapter~9 in \cite{BREZIS}.

\begin{Lemma} \label{lemma1}
(a) For all $u,v,w \in E$, we have
\begin{eqnarray*}
& &
a_1(u,v,w)=-a_1(u,w,v), \\
& &
|a_1(u,v,w)|\leq C\|u\|_{L^2(\Omega)^2}^{1/2} \|u\|_E^{1/2} \|v\|_E \|w\|_{L^2(\Omega)^2}^{1/2} \|w\|_{E}^{1/2} \ \ \mbox{with} \ C>0, \\
& &
a_1(u,v,v)= 0.
\end{eqnarray*}
(b) For all $u\in E$, $\eta,\zeta \in V$, we have
\begin{eqnarray*}
& &
b_1(u,\eta,\zeta)=-b_1(u,\zeta,\eta),\\
& &
|b_1(u,\eta,\zeta)|\le C\|u\|_{L^2(\Omega)^2}^{1/2} \|u\|_E^{1/2} \|\eta\|_V \|\zeta\|_{L^2(\Omega)}^{1/2} \|\zeta\|_V^{1/2}\ \ \mbox{with} \ C>0, \\
& &
b_1(u,\eta,\eta)= 0.
\end{eqnarray*}
\end{Lemma}

Finally, we recall the Green formula and the Aubin--Lions compactness lemma, 
which can be found in Theorem 2.25 in~\cite{MOSBOOK} and Corollary 4 in \cite{SIMON}, respectively.

\begin{Theorem}\label{green2}
Let $\Omega$ be a bounded domain in $\mathbb{R}^2$ with Lipschitz boundary $\Gamma$. Then for all $v\in H^1(\Omega)^2$ and $\sigma\in C^1(\bar{\Omega};\mathbb{S}^2)$ 
the following formula holds
\begin{equation}
\int_\Omega \sigma \colon \varepsilon (v)\,dx
+ \int_\Omega {\rm Div}\,\sigma \cdot v\,dx
= \int_\Gamma \sigma \nu \cdot v \, d\Gamma,
\end{equation}
where $\mathbb{S}^2$ denotes the linear space 
of second order symmetric tensors on $\mathbb{R}^2$ and $\sigma \colon \tau = \sigma_{ij}\tau_{ij}$.

\end{Theorem}

\begin{Lemma} \label{AubinLions}
Let $X$, $Y$, $Z$ be reflexive Banach spaces and $X\subset Y \subset Z$ continously with compact embedding $X\to Y$. Let $0 < T < \infty$. Then the space $\{u\in L^p(0,T;X)\mid {\dot u}\in L^q(0,T;Z)\}$ is compactly embedded into $L^p(0,T;Y)$ for $p$, $q\in (1,\infty)$.
\end{Lemma}

Throughout the paper we denote by $C$ a generic constant that can change value from line to line.

\section{Problem statement}\label{Statement}

Let $\Omega$ be a bounded domain in $\mathbb{R}^2$
with a regular boundary $\Gamma$ consisting of two nonempty sets  
$\Gamma_0$ and $\Gamma_1$. For a fixed and finite $T > 0$, 
consider the following Cauchy problem for 
nonstationary Boussinesq equations
\begin{eqnarray}
{\dot u} -\alpha \Delta u + (u\cdot \nabla) u + \nabla p &=& F(\theta) \ \ \mbox{in} \ \ \Omega \times (0,T) 
\label{NS} \\[2mm]
\mbox{div} \, u&=&0 \ \ \mbox{in} \ \ \Omega \times (0,T) \label{NS2} \\[2mm]
{\dot \theta} - \div (k(\theta) \nabla \theta) + u\cdot \nabla\theta & = & h \ \ \mbox{in} \ \ \Omega \times (0,T) \label{HEAT} \\[2mm]
u(0)=u_0, \ \theta(0) = \theta_0 & &\ \ \mbox{in} \ \ \Omega. \label{NS3}
\end{eqnarray}

\noindent 
The system (\ref{NS})--(\ref{NS3}) describes the incompressible viscous fluid flow 
in the domain $\Omega$, where $u \colon \Omega \times (0,T) \to \real^2$ 
denotes the fluid velocity, $\theta\colon \Omega\times(0,T)\to \real$ is the temperature, $F \colon \real \to \real^2$ 
is an external force vector field depending on $\theta$, $p \colon \Omega \times (0,T) \to \real$ is the pressure, $\alpha >0$ is the kinematic viscosity coefficient of the fluid and $k$ is the heat conductivity function. For the sake of simplicity, we investigate the isotropic case. The reader can easily generalize these results in anisotropic situation.
The divergence free condition (\ref{NS2}) states that the motion is incompressible. 

We supplement the system (\ref{NS})--(\ref{NS3}) with the following boundary 
conditions. We impose the adhesive boundary condition on part $\Gamma_0$, 
i.e.,
\begin{equation}
u=0 \ \mbox{on} \ \Gamma_0 \times (0, T) \label{zero}.
\end{equation}
On part $\Gamma_1$ we consider the nonmonotone friction law
\begin{equation}\label{SLIP}
u_\nu =0, \ \ \ -\sigma_{\tau} \in \partial j(u_\tau) \ \ \mbox{on} \ \ \Gamma_1 \times (0, T)
\end{equation}
which is also called the slip boundary condition.
Concerning the temperature, we assume that it is prescribed on $\Gamma_0$, i.e.
\begin{equation}
\theta = 0 \ \mbox{on} \ \Gamma_0 \times(0,T),
\end{equation}
and the heat flux through $\Gamma_1$ satisfies a nonmonotone law of the type
\begin{equation}\label{FLUX}
-k(\theta)\frac{\partial\theta}{\partial \nu} \in \partial j_1(\theta) \ \mbox{on} \ \Gamma_1 \times (0,T).
\end{equation}
In conditions (\ref{SLIP}) and (\ref{FLUX}), the functions 
$j \colon \Gamma_1\times (0,T)\times \mathbb{R}^2 \to \mathbb{R}$
and 
$j_1 \colon \Gamma_1\times (0,T)\times \mathbb{R} \to \mathbb{R}$
are assumed to be locally Lipschitz with respect to their last variable, 
and $\partial j$, $\partial j_1$ denote their Clarke 
subdifferentials. 
Here $\sigma$ stands for the standard stress tensor for 
incompressible fluid which is given by 
\begin{equation}
\sigma = - pI + 2 \alpha \varepsilon(u) \ \mbox{in} \ \Omega\times(0,T),
\label{stress}
\end{equation}
\noindent 
where $I$ is the identity matrix and $\varepsilon(u)=(\varepsilon_{ij}(u))$,  
$\varepsilon_{ij}(u) =\frac{1}{2}\left(\frac{\partial u_i}{\partial x_j} + \frac{\partial u_j}{\partial x_i}\right)$ is the strain tensor, $i, j = 1, 2$.

\medskip

We need the following 
hypotheses.

\smallskip
\smallskip

\noindent
${\underline{H(j)}}:$
$j \colon \Gamma_1\times \mathbb{R}^2 \to \mathbb{R}$ is such that
\begin{itemize}
\item[(a)]
$j(\cdot,\xi)$ is measurable for all $\xi \in \mathbb{R}^2$ 
and $j(\cdot,0) \in L^1(\Gamma_1)$.
\item[(b)]
$j(x, \cdot)$ is locally Lipschitz for a.e. $x \in \Gamma_1$.
\item[(c)]
$\| \eta \| \le c_0(1+\| \xi \|)$ for all $\xi \in \mathbb{R}^2$, $\eta \in \partial j(x,\xi)$, 
a.e. $x \in \Gamma_1$ with $c_0 > 0$.
\item[(d)]
$(\zeta_1 - \zeta_2) \cdot (\xi_1 - \xi_2) \ge -
m_1 \| \xi_1 - \xi_2 \|^2_{\mathbb{R}^2}$ for all 
$\zeta_i \in \partial j (x, \xi_i)$, $\xi_i \in \mathbb{R}^2$, $i =1$, $2$, 
a.e. $x \in \Gamma_1$ with $m \ge 0$.
\end{itemize}

\smallskip

\noindent
${\underline{H(j_1)}}:$
$j_1 \colon \Gamma_1\times \mathbb{R} \to \mathbb{R}$ is such that
\begin{itemize}
\item[(a)]
$j_1(\cdot,r)$ is measurable for all $r \in \mathbb{R}$ 
and $j_1 (\cdot,0) \in L^1(\Gamma_1)$.
\item[(b)]
$j_1(x, \cdot)$ is locally Lipschitz for a.e. $x\in \Gamma_1$.
\item[(c)]
$|s|\le c_1 (1 + | r |)$ for all $r \in \mathbb{R}$, $s \in \partial j(x,r)$, 
a.e. $x \in \Gamma_1$ with $c_1>0$.
\end{itemize}
\smallskip 
${\underline{H(F)}:}$ 
$F\colon \real \to \real^2$ is linear and continuous.
\medskip

\noindent 
${\underline{H(k)}:}$
$k\colon \mathbb{R} \to \real$ is bounded, Lipschitz continuous and  $k(r) > \delta$ for all $r\in \real$ with~$\delta>0$.

\medskip

\noindent
${\underline{H_0}:}$
$u_0\in E, \theta_0\in V$, $g\in L^2(0,T;V^*)$, $\alpha > \max \{ 2\sqrt{2} c_0, m \} \|\gamma_s\|^2, \ \delta > 2\sqrt{2}c_1 \|\gamma\|^2$. 

\medskip

In the following sections we will study a system of parabolic hemivariational inequalities which is a weak formulation of problem (\ref{NS})--(\ref{FLUX}).
With a slight abuse of notation, we will denote an operator and the Nemytskii operator associated to it by the same letter.

\section{Variational formulation}\label{SECSLIP}

In this section we provide the variational formulation of problem
(\ref{NS})--(\ref{FLUX}) and deliver a result on its solvability.

Assume that $u$, $p$ and $\theta$ are sufficiently smooth functions 
which solve (\ref{NS})--(\ref{FLUX}).
Let $v \in E$. 
By the Green formula of Theorem~\ref{green2}
applied to the relation (\ref{stress})
and by the incompressibility condition (\ref{NS2}), 
we have 
$$
\int_\Omega p\, ({\rm div} \, v) \, dx + 
2 \int_{\Omega} \varepsilon(u) \colon \varepsilon(v) \, dx + 
\int_{\Omega} (-\nabla p
\medskip + \alpha\Delta u) v \,dx  = 
\int_{\Gamma_1} \sigma \nu \cdot v \, d\Gamma.
$$ 
Hence
\begin{equation*}\label{green3}
\int_{\Omega}(-\alpha \Delta u + \nabla p ) v\, dx  - \int_{\Gamma_1} \sigma \nu \cdot v \,d\Gamma 
= a_0(u,v) + c(v,p).
\end{equation*}

\noindent 
Next, using equation (\ref{NS}) 
and definitions of operators $A_0$ and $A_1$, we find
\begin{equation}\label{MM1}
\left\langle {\dot u}(t) + A_0 u(t) + A_1 u(t), 
v \right\rangle_{E^*\times E} - 
\int_{\Gamma_1} \sigma\nu \cdot v \,d\Gamma = 
\langle F(\theta(t)), v \rangle_{{E}^*\times{E} }
\end{equation}
for all $v \in E$, 
a.e. $t\in(0,T)$. 
Since $v \in E$, by the orthogonality relation 
$\sigma \nu \cdot v = \sigma_\nu v_\nu + \sigma_\tau \cdot v_\tau$ 
on $\Gamma_1 \times (0, T)$ and conditions (\ref{SLIP}),
we get 
\begin{equation}\label{MM2}
- \int_{\Gamma_1} \sigma\nu \cdot v \, d\Gamma =
- \int_{\Gamma_1} \sigma_\tau \cdot v_\tau \, d\Gamma
\le 
\int_{\Gamma_1} j^0(u_\tau(t);v_\tau) \,d\Gamma
\end{equation}

\noindent 
for all $v\in E$, a.e. $t \in (0, T)$. Using again Green's formula, we have for $\zeta\in V$
\begin{align}
&
-\int_{ \Omega} \div(k(\theta)\nabla \theta) \zeta \, dx = \int_{\Omega} k(\theta)\nabla \theta \cdot \nabla \zeta \, dx - \int_{ \Gamma_1} k(\theta)\nabla \theta\cdot  \nu \ \zeta \, d\Gamma.
\end{align}
The relation (\ref{FLUX}) implies
\begin{equation}
- \int_{ \Gamma_1} k(\theta)\nabla \theta\cdot  \nu  \ \zeta \, d\Gamma \le \int_{\Gamma_1} j_1^0(x,\theta(t);\zeta) \, d\Gamma
\end{equation}
for a.e. $t\in (0,T)$.
Summarizing, we arrive at the following system of inequalities which represents the variational formulation of problem (\ref{NS})--(\ref{FLUX}).
\medskip

\begin{Problem}\label{Problem1}
 Find $u \in \mathbb{E}$ and $\theta \in \mathcal{W}$ such that
\begin{eqnarray}
&& \nonumber
\left\langle {\dot u}(t) + A_0 u(t) + A_1 u(t),v\right\rangle_{E^*\times E} + 
\int_{\Gamma_1} j^0(u_\tau(t);v_\tau) \,d\Gamma \ge \langle F(\theta(t)),v\rangle_{E^*\times E}  \\[2mm] 
&&
\hspace{7.0cm} \ \ \mbox{for all} \ \ v\in E, \ \mbox{a.e.} \ t\in(0,T), \label{NST} \\ 
&& \nonumber
\langle \dot{\theta}(t) + B_0 (\theta(t),\theta(t)) + B_1(u(t),\theta(t)),\zeta\rangle_{V^*\times V} + \int_{\Gamma_1} j_1^0(\theta(t);\zeta) \,d\Gamma \ge \langle g(t),\zeta \rangle_{V^*\times V} \\[2mm]
&&
\hspace{7.0cm} \ \ \mbox{for all} \ \ \zeta\in V, \ \mbox{a.e.} \ t\in(0,T), \label{NST1}\\
&&
u(0)=u_0, \ \theta(0)=\theta_0. \nonumber
\end{eqnarray}
\end{Problem}

\begin{Theorem}\label{t1}
Under hypotheses $H(j), H(j_1), H(F), H_0$, $H(k)$, 
Problem~\ref{Problem1} has at least one solution.
\end{Theorem}
{\bf Proof.} 
Proof of the theorem will be done in a few steps. 

{\bf Step 1.} In order to show the existence of solution we associate with Problem~\ref{Problem1} an operator evolution inclusion. To this end we define the functional 
$J\colon L^2(\Gamma_1)^2 \to \real$ by 
\begin{equation*}
J( u) = \int_{\Gamma_1} j(x, u_\tau(x))\, d\Gamma 
\ \ \mbox{for} \ u \in L^2(\Gamma_1)^2. 
\end{equation*}
Under hypothesis $H(j)$, the functional $J$ is locally Lipschitz and satisfies the following inequality (cf. Lemma 3 in \cite{MIGUNI})
\begin{equation}\label{propJ}
J^0(u;v)\le \int_{\Gamma_1} j^0 (x,u_\tau(x);v_\tau(x)) \, d\Gamma \quad \mbox{for all} \ u,v\in L^2(\Gamma_1)^2,
\end{equation}
where $J^0(u;v)$ and $j^0(x,u;v)$ denote the generalized directional derivative of $J$ and $j(x,\cdot)$, respectively. We also define the functional $J_1\colon L^2(\Gamma_1)\to \real$ by
\[
J_1(\zeta)=\int_{\Gamma_1} j_1(x,\zeta(x)) \, d\Gamma \ \ \mbox{for} \ \zeta \in L^2(\Gamma_1).
\]
Then, by $H(j_1)$ the functional $J_1$ is locally Lipschitz and enjoys the property
\begin{equation}\label{propJ}
J_1^0(\theta;\zeta)\le \int_{\Gamma_1} j_1^0 (x,\theta(x);\zeta(x)) \, d\Gamma \quad \mbox{for all} \ \theta,\zeta\in L^2(\Gamma_1).
\end{equation}

\noindent 
Consider the following system of inclusions associated with Problem~\ref{Problem1}: 
find $u\in \mathbb{E}$ and $\theta\in \mathcal{W}$ such that
\begin{eqnarray}
&& \label{Inc11}
\hspace{-1.0cm} {\dot u}(t) + A_0 u(t) + A_1 u(t) + \gamma_s^*\partial J(\gamma_s u (t)) \ni F(\theta(t))
\ \ \mbox{for a.e.} \ \ t \in (0,T), \\ [2mm]
&& \label{Inc111}
\hspace{-1.0cm}{\dot \theta}(t) + B_0 (\theta(t),\theta(t)) + B_1(u(t),\theta(t)) + \gamma^* \partial J_1(\gamma \theta(t)) \ni g(t) \ \ \mbox{for a.e.} \ \ t \in (0,T), \\ [2mm]
&&
\hspace{-1.0cm}u(0) = u_0, \ \theta(0)=\theta_0. \label{Inc12}
\end{eqnarray}
A solution $(u,\theta)\in\mathbb{E}\times\mathcal{W}$ of (\ref{Inc11})--(\ref{Inc12}) is understood is the sense that there exist $\xi\in L^2(0,T;L^2(\Gamma_1)^2)$ and $\xi_1\in L^2(0,T;L^2(\Gamma_1))$ such that
\begin{eqnarray}
&& \label{Inc1}
{\dot u}(t) + A_0 u(t) + A_1 u(t) + \gamma_s^* \xi(t) = F(\theta(t))
\ \ \mbox{for a.e.} \ \ t \in (0,T), \\[2mm]
&& 
\label{Inc3}
\xi(t) \in \partial J( \gamma_s u (t)) \ \ \mbox{for a.e.} \ \ t \in (0,T), \\[2mm]
&&  \label{Inc2}
{\dot \theta}(t) + B_0 (\theta(t),\theta(t)) + B_1(u(t),\theta(t)) + \gamma^* \xi_1(t) = g(t) \ \ \mbox{for a.e.} \ \ t \in (0,T),\\[2mm]
&& \label{Inc4}
\xi_1(t) \in \partial J_1(\gamma \theta(t))\ \ \mbox{for a.e.} \ \ t \in (0,T), \\[2mm]
&& \label{Inc5}
u(0) = u_0, \ \theta(0)=\theta_0. 
\end{eqnarray}

\noindent 
We observe that every solution to (\ref{Inc11})--(\ref{Inc12}) 
is also a solution to Problem~\ref{Problem1}. 
Therefore, in order to complete the proof, it is enough to establish 
the existence of a solution to problem (\ref{Inc11})--(\ref{Inc12}).

{\bf Step 2.} In this step, we introduce an auxiliary problem to (\ref{Inc1})--(\ref{Inc5}). To this end, we define spaces $U=V\cap W^{1,4}(\Omega)$, $\mathcal{U} = L^4(0,T;U)$, and the operator $G\colon U\to U^*$ by
\begin{equation}
\langle Gu,v \rangle_{U^*\times U} = \int_{\Omega} \|\nabla u\|_{\real^2}^2 \nabla u\cdot \nabla v\, dx \ \ \mbox{for all} \ \  u,v\in U. \label{PLAPL}
\end{equation}
 In the following we use the notation: for a function $g\colon [0,T]\to X$  defined everywhere on $[0,T]$, where $X$ is a reflexive Banach space, we write
\[
g_h(t)= \begin{cases}
g(t-h), & t> h\\
g(0), & t\in[0,h].
\end{cases}
\]
for $t\in (0,T)$. We observe that 
\begin{equation}
\|g_h\|_{L^2(0,T;X)}^2\le h \|g(0)\|_X^2 + \|g\|_{L^2(0,T;X)}^2.\label{nierow}
\end{equation}
\noindent
Indeed, we have
\begin{align*}
&
\int_0^T \|g_h(t)\|_X^2 \, dt  = \int_0^h \|g(0)\|_X^2\, dt + \int_h^T \|g(t-h)\|_X^2 \, dt = \\
&
=h\|g(0)\|_X^2 + \int_0^{T-h} \|g(s)\|_X^2 \, ds \le h \|g(0)\|_X^2 + \|g\|_{L^2(0,T;X)}^2.
\end{align*}

Fix $h>0, h\in (0,T)$. We introduce the regularized and time retarded problem.

\begin{Problem}\label{ProblemAp}
Find $u^h \in \mathbb{E},$ $\theta^h \in \mathcal{U}$ with $\dot{\theta}^h \in \mathcal{U}^*$ such that
\begin{eqnarray}
&& \label{Apr1}
\dot{u}^h + A_0(u^h) + A_1(u^h,u^h) + \gamma_s^* \xi^h = F(\theta_h^h)\ \mbox{in} \ \mathcal{E}^*, \\ [2mm]
&& \label{Apr2}
\xi^h(t)\in \partial J(u^h(t)) \ \mbox{for a.e.} \ t\in (0,T),\\ [2mm]
&& \label{Apr3}
u^h(0)=u_0, \\[2mm]
&& \label{Apr4}
 \dot{\theta}^h + B_0(\theta_h^h,\theta^h) + B_1(u_h^h,\theta^h) + hG\theta^h + \gamma^*\xi_1 = g \ \mbox{in} \ \mathcal{U}^*, \\ [2mm]
&& \label{Apr4a}
\xi_1^h(t) \in \partial J_1(\theta^h(t)) \ \mbox{for a.e.} \ t\in (0,T), \\ [2mm]
&& \label{Apr5}
\theta^h(0)= \theta_0.
\end{eqnarray}
\end{Problem}

The method used here to obtain Problem~\ref{ProblemAp} is called time--retardation (cf.  \cite{KUTSHIL}). The idea is to divide the time interval into finite number of intervals of length $h$ and do the backward translation in time. We then observe that on any such interval all elements with subscript $h$ are known. This allows to treat the two problems (\ref{Apr1})--(\ref{Apr3}) and (\ref{Apr4})--(\ref{Apr5}) separately and show existence of solution for each of them indepenently. 
The role of the operator $G$ given by (\ref{PLAPL}) and the space $U$ 
is to consider a regularized problem to (\ref{Inc2})--(\ref{Inc5}) and use an abstract result on the existence of solution, cf. Theorem 5 of~\cite{MIGPARA}, to problem (\ref{Apr4})--(\ref{Apr5}).

{\bf Step 3.} In this step, we show that under assumptions of the theorem, there exists a solution to Problem~\ref{ProblemAp}.

First, we observe that there exists a solution to (\ref{Apr1})-(\ref{Apr5}) on interval $[0,h]$. Indeed, on $[0,h]$, the functions $u_h^h$ and $\theta_h^h$ are given, since we have $\theta_h^h(t)=\theta_0$,  $u_h^h(t)=u_0$ for all $t\in [0,h]$. Therefore, on $[0,h]$ we can solve (\ref{Apr1})--(\ref{Apr3}) and (\ref{Apr4})--(\ref{Apr5}) independently. For (\ref{Apr1})--(\ref{Apr3}), we use the Galerkin method and for the moment we skip the superscript $h$, to restore it later. 

In order to solve (\ref{Apr1})--(\ref{Apr3}) we formulate the regularized problem as follows. 
Let $\rho\in C_0^{\infty}(\mathbb{R}^2)$ be the mollifier such that 
$\rho \ge 0$ on $\mathbb{R}^2$, 
$\mbox{supp}\,\rho \subset [-1,1]^2$ and $\int_{\mathbb{R}^2} \rho \, dx =1$. 
We define $\rho_m(x) = m^2 \rho(mx)$ for $m \in \mathbb{N}$.
Then $\mbox{supp}\, \rho_m \subset [-\frac{1}{m}, \frac{1}{m}]$ 
for all $m \in \mathbb{N}$. 
Consider functions 
$j_{m} \colon \Gamma_1\times\mathbb{R}^2 \to \mathbb{R}$ 
defined by
$$
j_{ m}(x,\xi) = \int_{supp\,\rho_n} \rho_m(z) j (x,\xi-z)\,dz 
\quad \mbox{for} \ (x,\xi) \in\Gamma_1\times\mathbb{R}^2.
$$
We observe that 
$j_{ m}(x,\cdot)\in C^\infty (\mathbb{R}^2)$ for all $x\in \Gamma_1$, 
so $\partial j_{m}(x,\xi)$ reduces to a single element, 
and we write 
$\partial j_{ m}(x,\xi(t)) = \{ D_u j_{ m}(x,\xi(t))\}$ for all $\xi(t)\in E$, where $D_u j_m$ represents the derivative of $j_m(x,\cdot)$. Moreover, it is easy to observe that $j_m$ satisfies the  growth condition $H(j)(c)$.

\
\noindent 
Using the fact that $E$ is a separable Banach space, 
we denote by $\{\varphi_1,\varphi_2,\ldots\}$ a basis of $E$. 
We also define $E^m = {\rm span}\, \{\varphi_1,\ldots,\varphi_m\}$ for $m\in \mathbb{N}$. Let $u_{0m}\in E^m$ for $m\ge 1$ be such that $u_{0m}\to u_0$ in $H$ as $m\to\infty$.

For a fixed $h>0$ we consider following regularized system of equations is finite dimensional space, corresponding to (\ref{Apr1})--(\ref{Apr3}).

\begin{Problem}\label{AUX1} 
Find $u_m\in L^2(0,h;E^m)$ such that $\dot{u} \in L^2(0,h;E^m)$ and
\begin{eqnarray}
&& \label{nr0}
\hspace{-1.0cm}
\langle {\dot u}_m(t) + A_0 u_m(t)+ A_1u_m(t),v_m \rangle_{E^*\times E} + 
(D_u j_{ m}(\gamma_s u_{m\tau}(t)),\gamma_s v_{m\tau})_{L^2(\Gamma_1)^2}  = 
\nonumber \\ [2mm]
&&
\hspace{1.0cm} 
= \langle F(\theta_h(t)),v_m \rangle _{E^*\times E} \ \ \mbox{for all}  \ v_m\in E^m , \ \ \mbox{a.e.} \ t\in(0,h), \label{fin}\\
\nonumber
&& \hspace{-1.0cm} u_m(0)= u_{0m}, \label{disrete}
\end{eqnarray}
\end{Problem}
\noindent
where $\theta_h(t)=\theta_0$ for all $t\in [0,h]$. We show that Problem~\ref{AUX1} has a solution. 
Substituting $u_m(t)=\sum_{k=1}^m c_{km}(t)\, \varphi_k$ in \eqref{fin} 
gives a system of first order ordinary differential equations 
for the coefficients $c_{km}\in C^1(0,h)$. Its solvability follows from the Carath\'eodory theorem. 

Now we show the a priori estimates for Problem~\ref{AUX1}. 
To this end, choose $u_m(t)$ as test function in \eqref{fin}.
Using the Young inequality and coercivity of operator $A_0$, we have
\begin{eqnarray}
&& \nonumber
\frac{1}{2}\frac{d}{dt}\|u_m(t)\|_H^2 + \alpha \|u_m(t)\|_E^2 + (D_u j_{ m}(\gamma_s u_{m\tau}(t)), \gamma_s u_{m\tau}(t))_{L^2(\Gamma_1)^2} \le \\ [2mm] 
&& \nonumber
\hspace{4.0cm} \leq \frac{\alpha}{2}\|u_m(t)\|_{E}^2 + \frac{2}{\alpha} \|F\|_{\mathcal{L}({L^2(\Omega);E^*})} \|\theta_h\|_{L^2(\Omega)}^2
\end{eqnarray}
for a.e. $t\in (0,h)$. Integrating over $(0,r)$, for $r\in (0,h)$, we obtain
\begin{eqnarray}
& &\nonumber
\hspace{3.0cm}
\frac{1}{2}\|u_m(r)\|_{H}^2 - \frac{1}{2}\|u_{0m}\|_{H}^2 + \frac{\alpha}{2} \int_0^r \|u_m(s)\|_{E}^2 \,ds  +\\[2mm] 
& &
\hspace{-1.0cm} + \int_0^r(D_u j_{ m}(\gamma_s  u_{m\tau}(s)),\gamma_s u_{m\tau}(s))_{L^2(\Gamma_1)^2}\, ds \leq 
\frac{2}{\alpha} \|F\|_{\mathcal{L}(L^2(\Omega);E^*)}  \|\theta_h\|_{L^2(0,h;L^2(\Omega))}^2 \ \  \label{nr1}
\end{eqnarray}
for a.e. $r\in (0,h)$. Exploiting the growth condition of $j_m$, following the proof of Theorem 1 in \cite{OCHAL1}, we get
\begin{eqnarray*}
 & &
\|D_uj_{ m}(\gamma_s u_{m\tau}(t))\|_{L^2(\Gamma_1)^2}^2 \leq 2c_0^2 \int_{\Gamma_1} (1+\|u_m(x,t)\|_{\mathbb{R}^2}^2) \,d\Gamma \leq \\ [2mm]
& &
\hspace{4.0cm} \leq 2c_0^2 m(\Gamma_1) + 2c_0^2 \|\gamma_s\|^2\|u_m(t)\|_{E}^2
\end{eqnarray*}
\noindent
for a.e. $t\in (0,r)$.
Hence 
\begin{equation}\
\|D_uj_{ m}(\gamma_s u_{m\tau})\|_{L^2(0,r;L^2(\Gamma_1)^2)}\leq c_a + c_b\|u_m\|_{L^2(0,r;E)} \ \mbox{for all} ~r\in (0,h) \label{boundD}
\end{equation}
with $c_a=c_0\sqrt{2hm(\Gamma_1)}$ and $c_b=c_0\sqrt{2}\|\gamma_s\|$. Consequently, we have
\begin{eqnarray}
&&\nonumber
\left|\int_0^r\left(D_u j_{\tau m}(\gamma_s u_{m\tau}(s)),\gamma_s u_{m\tau}(s)\right)_{L^2(\Gamma_1)^2} \,ds\right| \le \\ [2mm]
&&
\leq (c_a +c_b\|u_m\|_{L^2(0,r;E)})\|\gamma_s\| \|u_m\|_{L^2(0,r;E)} \label{nr2}
\end{eqnarray}
for all $r\in(0,h)$. Therefore, from (\ref{nr1}), (\ref{nr2}) and hypothesis $H(j)$, we deduce that $\{u_m\}$ remains bounded in $L^2(0,h;E)\cap L^\infty(0,h;H)$. By definition (\ref{oper}) and Lemma 3.4 in \cite{TEMAN1}, we have
\begin{equation} \label{2d}
\|A_1(u_m)\|_{L^2(0,h;E^*)} \le C\|u_m\|_{L^\infty(0,h;H)} \|u_m\|_{L^2(0,h;E)}
\end{equation}
with $C>0$ independent of $m$. 
Up to a subsequence we may assume that
\begin{equation}\label{Galcon}
u_m \to u \ \mbox{weakly in} \ L^2(0,h;E).
\end{equation}
From (\ref{fin}), (\ref{2d}), (\ref{Galcon}) and boundedness of $A_0$, we find that $\dot{u}_m$ is bounded in $L^2(0,h;E^*)$. Thus, the sequence $\{\dot{u}_m\}$ is bounded in a reflexive Banach space $L^2(0,h;E^*)$, and therefore, we see that
\begin{equation} \label{conv1}
\dot{u}_m \to \dot{u} \ \mbox{weakly in} \ L^2(0,h;E^*)
\end{equation}
with $u\in L^2(0,h;E^*)$. Since by Lemma~\ref{AubinLions} the embedding $\mathbb{E}^h =\{v\in L^2(0,h;E) \mid v' \in L^2(0,h;E^*) \}\subset L^2(0,h;H)$ is compact, from (\ref{Galcon}) and (\ref{conv1}), we have 
\[
u_m \to u \ \mbox{in} \ L^2(0,h;H).
\]
By the compactness of the trace operator from $\mathbb{E}^h$ into $L^2(0,h;L^2(\Gamma_1)^2)$, where by a slight abuse of notation we denote again by $\gamma_s$ the Nemytskii coresponding to $\gamma_s$, it follows
\[
\gamma_s u_m \to \gamma_s u \ \mbox{in} \ L^2(0,h;L^2(\Gamma_1)^2)
\]
and subsequently, by passing to a subsequence, if necessary, we have
\begin{equation} \label{1}
(\gamma_s u_{m\tau})(r) \to (\gamma_s u_m)(r) \ \mbox{in} \ L^2(\Gamma_1)^2 \ \mbox{for a.e.} \ r\in (0,h).
\end{equation}
Next, applying Lemma~\ref{AubinLions} to the evolution triple of spaces $E\subset L^4(\Omega)^2 \subset E^*$, from (\ref{Galcon}) and (\ref{conv1}), we obtain
\[
u_m \to u \ \mbox{in} \ L^2(0,h;L^4(\Omega)^2).
\]
Hence, by Lemma 3.2 in \cite{TEMAN1}, we have
\[
A_1(u_m)\to A_1(u) \ \mbox{weakly in} \ L^2(0,h;E^*).
\]
Since $A_0\colon E\to E^*$ is a linear and continuous operator, so is its Nemytskii operator.
Therefore, we find that $A_0 u_m \to A_0 u$ weakly in $L^2(0,h;E^*)$.
On the other hand, by (\ref{boundD}), we may suppose that
\begin{equation} \label{2}
D_u j_{ m} (\cdot,\gamma_s u_{m\tau}(\cdot)) \to \eta \ \mbox{weakly in} \ L^2(0,h;L^2(\Gamma_1)^2)
\end{equation}
with $\eta \in L^2(0,h;L^2(\Gamma_1)^2).$
 Using convergences (\ref{1}) and (\ref{2})  and applying the Aubin--Cellina convergence theorem (see \cite{AUBIN}, Theorem 7.2.1) to the inclusion $
 D_u j_{ m}(\gamma_s u_{m\tau}(t)) \in \partial j_{m}(\gamma_s u_{m\tau}(t))$ for a.e. $t\in (0,h)$, 
  we get that $\eta(t) \in {\rm \bar{co}} \, \partial j (\gamma_s u_\tau(t)) = \partial j(\gamma_s u_\tau(t))$ for a.e. $t\in (0,h)$, where ${\rm {\bar{co}}}$ denotes the closure of the convex hull of a set.  
  Since the mapping $\mathbb{E}^h \ni w  \to w(0) \in H$ is linear and continuous, from (\ref{Galcon}) and (\ref{conv1}), we have $u_m(0) \to
u(0)$ weakly in $H$, which together with $u_{0m} \to u_0$ in $H$ entails $u(0) = u_0$.
Thus, we have proved that $u\in \mathbb{E}^h$ is a  solution to (\ref{Apr1})--(\ref{Apr3}).

Now we pass to the problem (\ref{Apr4})--(\ref{Apr5}).
Since the operator $B_0(\theta_h^h,\cdot) + B_1(u_h^h,\cdot) + hF\colon L^4(0,h;U) \to L^{4/3}(0,h;U^*)$ can be shown to be $L$--pseudomonotone, coercive and bounded, the existence of a solution $\theta^h\in \mathcal{U}$ with $\dot{\theta}^h\in \mathcal{U}^*$ to (\ref{Apr4})--(\ref{Apr5}) is guaranteed by Theorem 5 in \cite{MIGPARA}. Hence, we have solved the system (\ref{Apr1})--(\ref{Apr5}) on the interval~$[0,h]$. 

On the interval $[h,2h]$ functions $u_h^h$ and $\theta_h^h$ are again known, so we apply the method described above to find a solution on the interval $[h,2h]$. Observe that $$\{ u\in L^2(0,h;E) \mid \dot{u}\in L^2(0,h;E^*)\} \subset C(0,h;H)$$ and $$\{ \theta\in L^4(0,h;U) \mid \dot{\theta}\in L^{4/3}(0,h;U^*)\} \subset C(0,h;L^2(\Omega)),$$ therefore $u^h(h)$ and $\theta^h(h)$ obtained as solution on $[0,h]$ now make sense as intial conditions in $H$ and $L^2(\Omega)$, respectively, for the problem on $[h,2h]$. We continue this process to obtain a solution $u^h\in \mathbb{E}$ and $\theta^h\in \mathcal{U}$ with $ \dot{\theta}^h \in \mathcal{U}^*$ to Problem~\ref{ProblemAp} on the whole interval $(0,T)$.

{\bf Step 4.} We now show the a priori estimates for Problem~\ref{ProblemAp}. 
Multiply (\ref{Apr1}) by $u^h$ and integrate over $[0,t]$, $t\in (0,T)$, use the hypotheses $H_0$, $H(j)$ and properties of $A_0$, $A_1$ to get
\begin{eqnarray}\label{AP1}
\frac{1}{2}\|u^h(t)\|_H^2 + \left( \alpha - c\|\gamma_s\|^2 -\varepsilon \right)
 \int_0^t \|u^h(s)\|_E^2 \, ds \le C + C\int_0^t \|\theta_h^h(s)\|_{L^2(\Omega)}^2 \, ds,
\end{eqnarray}
where $\varepsilon >0$ is sufficiently small and $C$ is indepentent of $h$. Next, we multiply (\ref{Apr2}) by $\theta^h$, integrate over $[0,t]$, $t\in (0,T)$, use hypothesis $H_0$, $H(j_1)$, properties of operators $B_0$, $B_1$ and we find
\begin{eqnarray}\label{AP2}
\frac{1}{2} \|\theta^h(t)\|_{L^2(\Omega)}^2 + \left(\delta - c_1\|\gamma\|^2 -\varepsilon\right) \int_0^t \|\theta^h(s)\|_V^2 \,ds  + h\int_0^t \|\nabla \theta(s)\|_{L^2(\Omega)}^4\, ds\le C.
\end{eqnarray}
\noindent
We choose $\varepsilon$ small enough and adjust the constants. Putting (\ref{AP2}) in (\ref{AP1}) and using (\ref{nierow}), we have
\begin{align}
&\nonumber
\frac{1}{2}\|u^h(t)\|_H^2 + \int_0^t \|u^h(s)\|_E^2 \, ds + h \int_0^t \|\nabla \theta^h(s)\|_{L^2(\Omega)}^4\, ds+ \\[2mm]
&
+\frac{1}{2} |\theta^h(t)|_{L^2(\Omega)}^2 + \int_0^t \|\theta^h(s)\|_V^2 \,ds \le C,\label{apriori}
\end{align}
for $t\in (0,T)$, where $C>0$ is independent of $h$.  From (\ref{apriori}), we can easily see that $\dot{u}^h$ and $\dot{\theta}^h$ are bounded in $\mathcal{E}^*$ and $\mathcal{U}^*$, respectively.  Therefore, we conclude
\begin{align}
&\label{con1}
u^h \ \mbox{is bounded in} \ L^\infty(0,T;H)  \ \mbox{and} \ L^2(0,T;E)\\[2mm]
&\label{con2}
\dot{u}^h \ \mbox{is bounded in}\  L^{2}(0,T;E^*),\\[2mm]
&\label{con3}
\theta^h \ \mbox{is bounded in} \ L^\infty(0,T;L^2(\Omega))  \ \mbox{and} \ L^2(0,T;V),\\[2mm]
&\label{con4}
\dot{\theta}^h \ \mbox{is bounded in} \ L^{4/3}(0,T;U^*), \\[2mm]
&\label{con5}
\xi^h \ \mbox{is bounded in} \ L^2(0,T;L^2(\Gamma_1)^2), \\[2mm]
&\label{con6}
\xi_1^h \ \mbox{is bounded in} \ L^2(0,T;L^2(\Gamma_1)).
\end{align}
 From (\ref{con3}), (\ref{con4}) and  Lemma~\ref{AubinLions}, we get that 
 \begin{equation}\label{thetasol}
 \theta^h \to \theta \ \ \mbox{in} \ L^r (0,T;L^2(\Omega)),
 \end{equation}
where $\theta\in L^r(0,T;L^2(\Omega))$ with $r\ge 1$.
Using (\ref{nierow}), we have
\begin{eqnarray}
\nonumber
\|\theta_h^h - \theta\|_{L^2(0,T;L^2(\Omega))} &\le & \|\theta_h^h-\theta_h\|_{L^2(0,T;L^2(\Omega))} + \|\theta_h - \theta\|_{L^2(0,T;L^2(\Omega))} \le\\ [2mm]
\label{nierownosc}
&\le &\|\theta^h -\theta\|_{L^2(0,T;L^2(\Omega))} + \|\theta_h-\theta\|_{L^2(0,T;L^2(\Omega))}.
\end{eqnarray}
The first term on the right hand side of (\ref{nierownosc}) converges to zero from (\ref{thetasol}) and the second from the continuity of translations in $L^2$ (see \cite{KUTSHIL}, p. 325).

{\bf Step 5.} In this step we introduce a function $u$ and show that the sequence $\{u^h\}$ obtained in Step~3 converges to $u$ in $L^2(0,T;E)$.

Let $\theta\in L^2(0,T;L^2(\Omega))$ be the function obtained in (\ref{thetasol}). We define $u\in \mathbb{E}$ to be a solution to the following problem. Find $u\in \mathbb{E}$ such that
\begin{eqnarray}
&& \label{u1}
\dot{u} + A_0(u) + A_1(u) + \gamma_s^* \xi = F(\theta) \ \mbox{in} \ \mathcal{E}^*,\\[2mm]
&& \label{u2}
\xi(t)\in \partial J(u(t)) \ \mbox{for a.e.} \ t\in (0,T) ,\\[2mm]
&& \label{u3}
u(0)=u_0.
\end{eqnarray}
The existence of a solution to (\ref{u1})--(\ref{u3}) follows from the Galerkin method, see Step~3. 
Now, we show that 
\begin{equation}\label{uconv}
u^h \to u \ \ \mbox{in} \ \mathcal{E},
\end{equation}
 where $u^h\in\mathbb{E}$ is a solution to Problem~\ref{ProblemAp} and $u\in\mathbb{E}$ is a solution to (\ref{u1})--(\ref{u3}). 
 
To show (\ref{uconv}), we subtract (\ref{Apr1}) from (\ref{u1}), and integrate over $(0,t)$, where $t\in (0,T)$,  use $H(j)(d)$ and the coercivity of $A_0$ to get
\begin{eqnarray}
&\nonumber
\displaystyle \frac{1}{2} \|u^h(t)-u(t)\|_H^2 + \left(\alpha - m_1\|\gamma_s\|^2-\varepsilon\right) \int_0^t \|u^h(s)-u(s)\|_E^2 \,ds \le \\ [2mm]
&\label{pom2}
\displaystyle 
\ \ \ \le C\int_0^t \|\theta_h^h(s)-\theta(s)\|_{L^2(\Omega)}^2 \, ds + 
C\int_0^t \|u^h(s)\|_E^2 \|u^h(s)-u(s)\|_H^2 \, ds.
\end{eqnarray}
for $\varepsilon>0$ small and $C$ independent of $h$. Using the Gronwall inequality, we obtain
\begin{equation}\label{pom1}
\|u^h(s)-u(s)\|_H^2 \le \int_0^t \|\theta_h^h(s)-\theta(s)\|_{L^2(\Omega)}^2 \, ds \exp\left(\int_0^t\|u^h(s)\|_E^2 \, ds\right).
\end{equation}
Since $\|u^h\|_{\mathcal{E}} \le C$, we put (\ref{pom1}) into (\ref{pom2}) and using (\ref{nierownosc}), we obtain (\ref{uconv}). 

Observe, that similarly to (\ref{nierownosc}), using (\ref{uconv}) we can show  that
$
\|u_h^h - u\|_{L^2(0,T;E)} \to 0.
$
\noindent

{\bf Step 6.} In this part of the proof, we show the convergence of all elements in (\ref{Apr4})--(\ref{Apr5}). Then we pass to the limit in (\ref{Apr4})--(\ref{Apr5}) to show the existence of solution to Problem~\ref{Problem1}.
The following convergences hold
\begin{align}
&\label{zb0}
B_0(\theta_h^h,\theta^h) \to B_0(\theta,\theta) \ \mbox{in} \  \mathcal{V}^* \\[2mm]
&\label{zb1}
B_1(u_h^h,\theta^h) \to B_1(u,\theta)\  \mbox{weakly in} \ \mathcal{U}^* \\[2mm]
&\label{zb2}
hG\theta^h \to 0 \  \mbox{in} \ \mathcal{U}^* \\[2mm]
&\label{zb3}
\gamma^* \xi_1^h \to \gamma^* \xi_1  \  \mbox{weakly in} \  \mathcal{V}^*,
\end{align}
as $h\to 0$.
\noindent
To prove (\ref{zb0}), we observe that
\begin{align}\nonumber
&
\langle B_0(\theta_h^h,\theta^h) - B_0(\theta,\theta),\zeta\rangle_{\mathcal{V}^*\times \mathcal{V}} = \int_0^T \int_{ \Omega} k(\theta_h^h)\nabla (\theta^h-\theta)\cdot \nabla \zeta \, dx \, dt + \\
&\label{pom3}
+ \int_0^T \int_{ \Omega} (k(\theta_h^h)-k(\theta)) \nabla \theta \cdot \nabla \zeta \, dx \, dt \ \ \mbox{for}\  \zeta\in \mathcal{V}.
\end{align}
The first integral on the right hand side of (\ref{pom3}) converges to zero, by the boundedness of $k$ and convergence $\theta^h\to\theta$ weakly in $\mathcal{V}$, obtained from (\ref{con3}). For the second integral, we use Lipschitz continuity of $k$ and the Lebesgue diminated convergence theorem. Hence, (\ref{zb0}) follows. 

For the proof of (\ref{zb1}), we use the Vitali convergence theorem. We can easily deduce from (\ref{con1}) and (\ref{con3}), the pointwise convergence
\begin{equation}\label{VitaliP}
u_h^h \, \theta^h \, \nabla \zeta(x,t) \to u \, \theta \, \nabla \zeta(x,t) \ \ \mbox{for a.e.} \  (x,t) \in\Omega\times (0,T).
\end{equation}
We calculate
\begin{equation*}
\int_0^T \int_{\Omega} u_h^h \, \theta^h \, \nabla \zeta \, dx\, dt \le C\int_0^T \|u_h^h\|_E^{1/2} \|\theta\|_V^{1/2} \|\zeta\|_V \, dt 
\le C \|\zeta\|_{\mathcal{V}} \le C\|\zeta\|_{\mathcal{U}}.
\end{equation*}
\noindent
This estimate together with pointwise convergence (\ref{VitaliP}) allows to use the Vitali theorem and so the convergence (\ref{zb1}) holds. 

The convergece (\ref{zb2}) follows from (\ref{apriori}) and the estimate
\begin{align*}
&
\int_0^T \langle hG\theta^h(t), \zeta(t) \rangle_{U^*\times U} \, dt \le h^{1/4} \left(\int_0^T h\langle G\theta^h(t), \theta^h(t) \rangle_{U^*\times U} \, dt \right)^{3/4} \|\zeta\|_{L^4(0,T;U)} \le \\[2mm]
&
\hspace{5.0cm}\le h^{1/4}C  \|\zeta\|_{L^4(0,T;U)}.
\end{align*}

Finally, the convergence (\ref{zb3}) follows from (\ref{con6}) and Aubin--Celina Theorem (see Theorem~7.2.1 in~\cite{AUBIN}) applied to the inclusion $\xi_1^h(t) \in \partial J(\gamma \theta^h(t))$ for a.e. $t\in (0,T)$. Namely, by the compactness of the trace operator we have 
\begin{equation*} 
(\gamma \theta^h)(t) \to (\gamma \theta)(t) \ \mbox{in} \ L^2(\Gamma_1) \ \mbox{for a.e.} \ t\in (0,T),
\end{equation*}
and by (\ref{con6}) we have $\xi_1^h\to \eta$ weakly in $L^2(0,T;L^2(\Gamma_1))$ for some $\eta\in L^2(0,T;L^2(\Gamma_1))$. Using the Aubin--Celina Theorem, we have $\eta(t)\in\bar{conv} \partial J( \gamma \theta(t))$ for a.e. $t\in (0,T)$, which finishes the proof of (\ref{zb3}).

Now we show the convergence of the initial condition.
By the compactness of the embedding 
\begin{equation}
S=\{ \theta\in L^\infty(0,T;L^2(\Omega)) \mid \dot{\theta}\in L^{4/3}(0,T;U^*)\} \subset C(0,T;U^*),
\end{equation}
(see \cite{SIMON}, Corollary 4), we have $\theta\in C(0,T;U^*)$. Since the mapping $S\ni \theta \to \theta(0) \in U^*$ is linear and continuous, by (\ref{con3})--(\ref{con4}), we have $\theta^h(0)\to \theta(0)$ weakly in $U^*$. This together with $\theta_0^h \equiv \theta_0$ in $U^*$ implies $\theta(0)=\theta_0$.

 Hence, passing to the limit in (\ref{Apr4})--(\ref{Apr5}), by (\ref{zb0})--(\ref{zb3}), we conclude that the following system has a solution $\theta\in \mathcal{U}$ with $\dot{\theta}\in \mathcal{U}^*$
 \begin{align}
 & \label{th1}
 \dot{\theta} + B_0(\theta,\theta) + B_1(u,\theta) + \gamma^*\xi_1 = g \ \ \mbox{in} \ \ \mathcal{U}^* \\[2mm]
 &\label{th2}
 \xi_1(t) \in \partial J_1(\theta(t)) \ \ \mbox{for a.e} \ \ t \in (0,T) \\[2mm]
 &\label{th3}
 \theta(0)=\theta_0,
 \end{align}
 where $u\in\mathbb{E}$ is a solution to (\ref{u1})--(\ref{u3}).
 To show existence of a solution to the original Problem~\ref{Problem1}, we need to show additionally that $\dot{\theta}$ is more regular, that is ${\dot \theta}\in L^2(0,T;V^*)$. It now follows easily from (\ref{con1})--(\ref{con4}), since
\begin{align*}
&
\|B_1(u,\theta)\|_{L^2(0,T;V^*)} =\sup_{\|\zeta\|_{L^2(0,T;V)}=1} \int_0^T\int_{ \Omega } u \ \theta \nabla \zeta \, dx \,dt\le \\[2mm]
&
\le C \sup_{\|\zeta\|_{L^2(0,T;V)}=1} \int_0^T \|u\|_E^{1/2} \|\theta\|_V^{1/2} \|\zeta\|_V \le C\sup_{\|\zeta\|_{L^2(0,T;V)}=1}  \|u\|_{\mathcal{E}} \|\theta\|_\mathcal{V} \|\zeta\|_\mathcal{V}
\end{align*}
for $u\in\mathbb{E}$, $\theta\in \mathcal{V}$, solutions to (\ref{u1})--(\ref{u3}) and (\ref{th1})--(\ref{th3}), with $C>0$.
Therefore, we have shown the existence of a solution to the following system: find $u\in \mathbb{E}$, $\theta\in \mathcal{W}$ such that
\begin{align}
&\label{fin1}
\dot{u} + A_0(u) + A_1(u,u) + \gamma_s^* \xi = F(\theta) \ \ \mbox{in} \  \ \mathcal{E}^* \\[2mm] 
&
\xi(t)\in \partial J(u(t)) \ \ \mbox{for a.e.} \ \ t\in (0,T) \\[2mm]
&
u(0)=u_0, \\[2mm]
&
\dot{\theta} + B_0(\theta,\theta) + B_1(u,\theta) + \gamma^*\xi_1 = g \ \ \mbox{in} \ \ \mathcal{V}^* \\
&
\xi_1(t) \in \partial J_1(\theta(t)) \ \ \mbox{for a.e} \ \ t \in (0,T) \\[2mm]
&\label{fin5}
\theta(0)=\theta_0.
\end{align}
\noindent
Finally, we observe that existence of a solution to (\ref{Inc1})--(\ref{Inc5}) is equivalent to the existence of solution to (\ref{fin1})--(\ref{fin5}). Since every solution to the problem (\ref{Inc1})--(\ref{Inc5}) is a solution to Problem~\ref{Problem1}, we have proved the thesis.
$\hfill{ \square}$

\begin{Remark}
In a standard way we can recover the pressure $p\in L^2(0,T;L^2(\Omega))$ in the original problem (\ref{NS})--(\ref{FLUX}). It follows from Proposition I.1.2 in \cite{TEMAN1} that $p(t)\in L^2(\Omega)$ for a.e. $t\in (0,T)$. 
\end{Remark}

\end{document}